\documentclass[intlimits,twoside,a4paper]{article}

\usepackage{amsmath,amssymb}
\usepackage{graphicx}
\usepackage{wrapfig}

\usepackage[T2A]{fontenc}
\usepackage[cp1251]{inputenc}

\usepackage[eqsecnum]{cmpj2}

\issue{2012}{15}{4}{43602}

\doinumber{10.5488/CMP.15.43602}

\title[Uranyl assocation: pH and temperature influence]
{Temperature and pH driven association in uranyl aqueous solutions}

\author{M. Druchok, M. Holovko}
\address{
Institute for Condensed Matter Physics, NAS of Ukraine,
1 Svientsitskii St., 79011 Lviv, Ukraine
}

\authorcopyright{M. Druchok, M. Holovko, 2012}

\date{Received July 3, 2012, in final form September 19, 2012}

\begin{document}

\maketitle

\begin{abstract}
An association behavior of uranyl ions in aqueous solutions is
explored. For this purpose a set of all-atom molecular dynamics
simulations is performed. During the simulation, the fractions of
uranyl ions involved in dimer and trimer formations were
monitored. To accompany the fraction statistics one also collected
distributions characterizing average times of the dimer and trimer
associates. Two factors effecting the uranyl association were
considered: temperature and pH. As one can expect, an
increase of the temperature decreases an uranyl capability of forming
the associates, thus lowering bound fractions/times and vice versa. The
effect of pH was modeled by adding H$^+$ or OH$^-$ ions
to a ``neutral'' solution. The addition of hydroxide ions OH$^-$
favors the formation of the associates, thus increasing bound times and fractions.
The extra H$^+$ ions in a solution produce an opposite effect, thus
lowering the uranyl association capability. We also made a structural analysis for all the observed
associates  to reveal the mutual
orientation of the uranyl ions.
\pacs 61.20.Ja, 61.20.Qg, 82.20.Wt, 82.30.Hk, 82.30.Nr
\end{abstract}

\section{Introduction}

For the last decades, the unresolved problem of  nuclear fuel wastes containing
actinides and other radionuclides contacting  with water urged intensive theoretical
and experimental studies. Comprehension of the association processes in such
systems can provide one with important data for chemical technology,
medicine, environmental ecology~\cite{Ahearne}. The actinides An in water can
easily form dioxides called actynils (AnO$_{2}$)$^{Z+}$. Both An${=}$O bond
lengths are usually 1.7--1.8~{\AA} and O${=}$An${=}$O angle is close to 180$^{\circ}$.
Actynil is usually hydrated by five water molecules, the so-called ligands,
located in an equatorial plane (normal to O${=}$An${=}$O axis) at the distances of
2.5--2.6~\AA. Such a finding was confirmed by quantum-chemical investigations.
In particular, in~\cite{Spencer,Tsushima00} it is shown that the number of
ligands of uranyl, neptunyl and plutonyl is equal to five. It was also found
that a hydrolysis reaction with one proton loss from one of five uranyl ligands
is energetically favorable. However, the quantum-chemical calculations cannot provide a scrupulous understanding of the role of the surroundings beyond the
hydrated-hydrolyzed complex, which is essential for a correct interpretation
of many structural, dynamic, and thermodynamic properties of these
complexes~\cite{Liu}. In \cite{Guilbaud93,Guilbaud96,Greathouse7},
the molecular dynamics simulations were carried out for aqueous solutions
of actinides using rigid TIP3P or SPC water models. Neither the deformation of
water molecules nor the hydrolysis effects in a hydration shell were possible
due to constrained rigid models of water molecules. No hydrolysis
evidence was also found in a recent investigation of uranyl hydration by
the {\it ab initio} molecular dynamics simulation~\cite{Nichols}.

In computer simulations, in order to explicitly treat the cation hydrolysis effect, water should be considered in the framework of a non-constrained
flexible model. In  our previous studies
\cite{hol9,Druchok10,hol11,hol12,NATO_uranyl}, water was treated within
a slightly modified version of central force model CF1~\cite{Nyberg,Duh15}.
It was found that with an increase of the cation-water interaction, some water
molecules in the cation hydration shell lost some protons. This effect was
treated as the hydrolysis of water caused by a high valency of the cation. In
\cite{Druchok16,NATO_uranyl}, the CF1 model for  water was  also used for
the investigation of the hydration structure of the uranyl ion UO$_{2}^{2+}$.
It is found that the uranyl hydration shell has bipyramydal pentacoordinated
structure with five ligands. It includes four water molecules and one hydroxide
OH$^-$ ion. Usually the cation hydrolysis process does not stop at the
creation of the hydrated-hydrolyzed complexes and proceeds to a condensation
reaction creating polynuclear ions~\cite{Ram17}.

Apparently, the tendency for the cation hydrolysis depends on the pH of
an aqueous solution. One can expect a hydrolysis effect in an alkaline
solution stronger than in an acidic one. Some previous results for the computer
modeling of the effect of pH on the cation hydrolysis of an uranyl ion
UO$_{2}^{2+}$ were presented in \cite{NATO_uranyl}. In this paper we
continue the investigation of the formation of the ionic associates between
uranyl ions. In particular, we study the effect of the temperature and pH
of an aqueous solution on the structure of the uranyl dimers and trimers.

\section{Model and method}

Three different solutions are considered. The first one is a solution
with 1600 water molecules and 16 uranyl ions. To mimic an acidic or
alkaline conditions, we add 100
H$^+$ or 100 OH$^-$ ions to the initial ``neutral'' solution, respectively. For simplicity we will
further refer to these solutions as alkaline, neutral, and acidic.

The model is similar to the one used in our study of the uranyl
hydration~\cite{Druchok16}. Central force model CF1 was engaged in water
description. For the uranyl-water interaction we took the potentials from
\cite{Greathouse7}. These potentials are of the ``1--12--6'' type:
\begin{equation}
E_{ij}=\frac {Z_i Z_j}{4\pi\epsilon_0 r}+\frac{A_i A_j}{r^{12}}-\frac{B_i B_j}{r^6}\,.
\end{equation}
The corresponding potential parameters are listed in table~\ref{tableU1}.
\begin{table}[h]
\caption{Potential parameters for uranyl-water interaction.}
\label{tableU1}
\begin{center}
\begin{tabular}{|l|c|c|c|}\hline
                & Z & $\vphantom{1^{1^1}}$ $A$~(kcal \AA$^{12}$/mol)$^{1/2}$ & $B$~(kcal \AA$^6$/mol)$^{1/2}$  \\ \hline\hline
O in H$_2$O     & --0.65966 & 793.322 & 25.010  \\
H               &  0.32983 &   0.1   &  0.0    \\
U               &  2.50    & 629.730 & 27.741  \\
O in UO$_2$     & --0.25    & 793.322 & 25.010  \\ \hline
\end{tabular}
\end{center}
\end{table}

In order to preserve the uranyl intramolecular geometry, additional
constraints are used for the U${=}$O bond length in the form
$E_{ij}\sim (r-1.75)^2$ and for the O${=}$U${=}$O angle in the form $E_{ij}\sim
(\theta-180)^2$. The distances are measured in~\AA, the angles -- in
degrees, the energies -- in~kcal/mol.

All the species were allowed to move freely across the MD cell. The cut-off
radius for the short-range interactions is chosen to be 15~\AA. The
long-range Coulomb interactions were taken into account by the Ewald
summation technique. One has to note that all the systems considered
are physically non-neutral but still can be effectively treated because the
charges are implicitly compensated by a neutralizing background in
the Ewald formulation. The pressure (1~bar) and the temperature (278~K or
318~K) were controlled by means of a Nose-Hoover barostat and thermostat in an
isotropic $NPT$ ensemble~\cite{Hayle,Melchionna}. The particles were placed
into a cubic box ($L_x=L_y=L_z\approx37$~\AA) with periodic boundary
conditions. The length of the production runs ranged from 7 to 10~ns. As
before~\cite{Druchok10,hol11,hol12,NATO_uranyl}, we used the velocity Verlet
algorithm with a time step 0.2~fs to integrate the classical equations of
motion.

For structural analysis, the radial distribution functions~(RDFs) are collected.
The angular distributions of uranyl-uranyl mutual orientation are also
accumulated during the simulations. To accompany structural details,
we have also collected distributions of lifetimes of the uranyl dimers and
trimers.

\section{Numerical results}

In this section we present the results of the simulations.
It includes the RDFs describing the uranyl-solution and uranyl-uranyl
correlation. As it is mentioned above, we also collected angular distributions.
Further we discuss distributions of lifetimes of the uranyl dimers and trimers.

\subsection{Radial distribution functions}

\begin{figure}[h]
\centerline{
\includegraphics[width=0.48\textwidth]{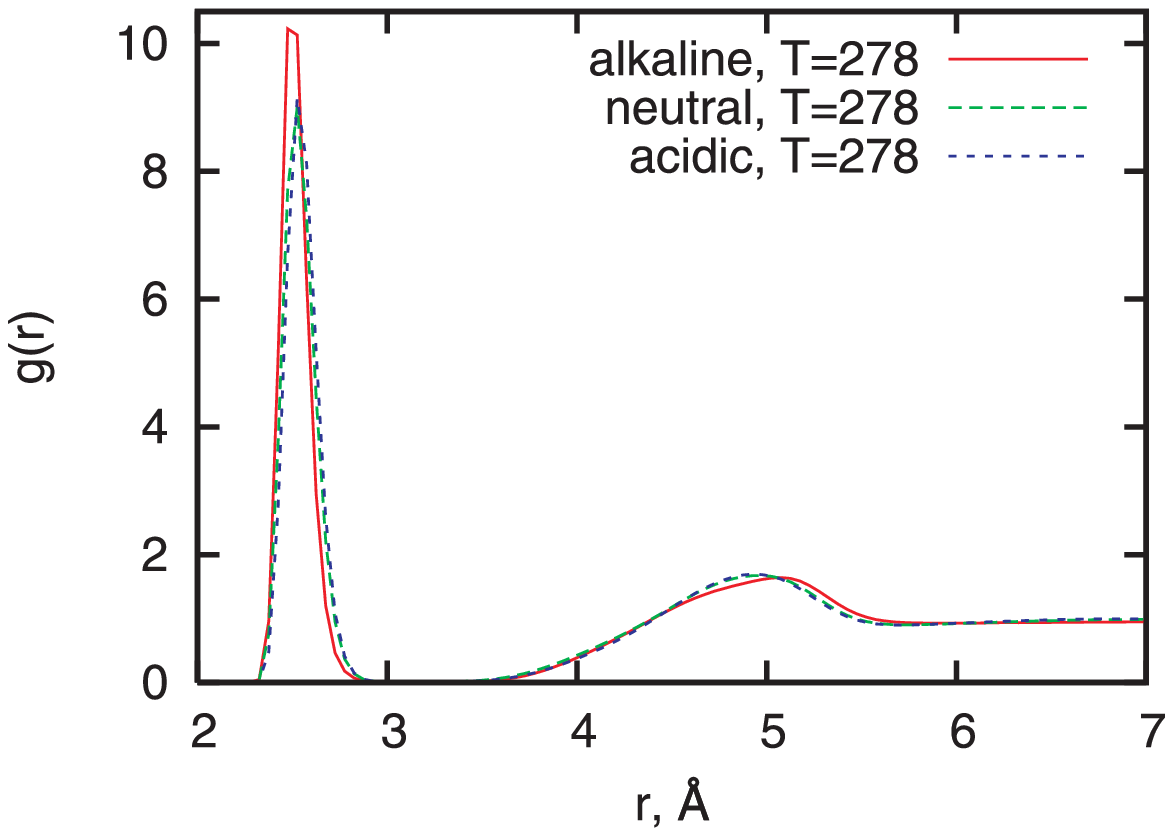}
\hfill
\includegraphics[width=0.48\textwidth]{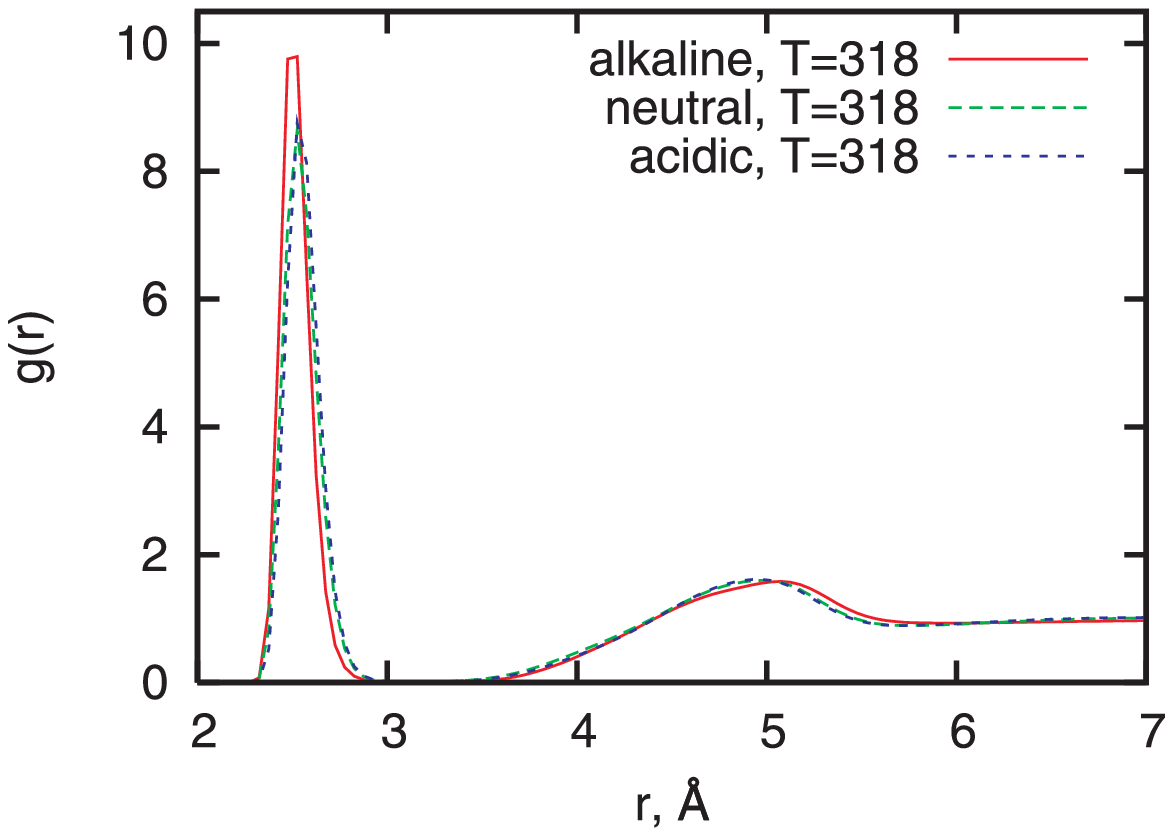}
}
\caption{(Color online) The uranium-oxygen~(of water and hydroxide ions) RDFs for
$T=278$~K~(left) and 318~K~(right). Results for the alkaline solution
are shown by solid red lines, neutral -- by dashed green lines, and acidic --
by dotted blue lines.}
\label{Fig:o-u}
\end{figure}

\begin{figure}[h]
\centerline{
\includegraphics[width=0.48\textwidth]{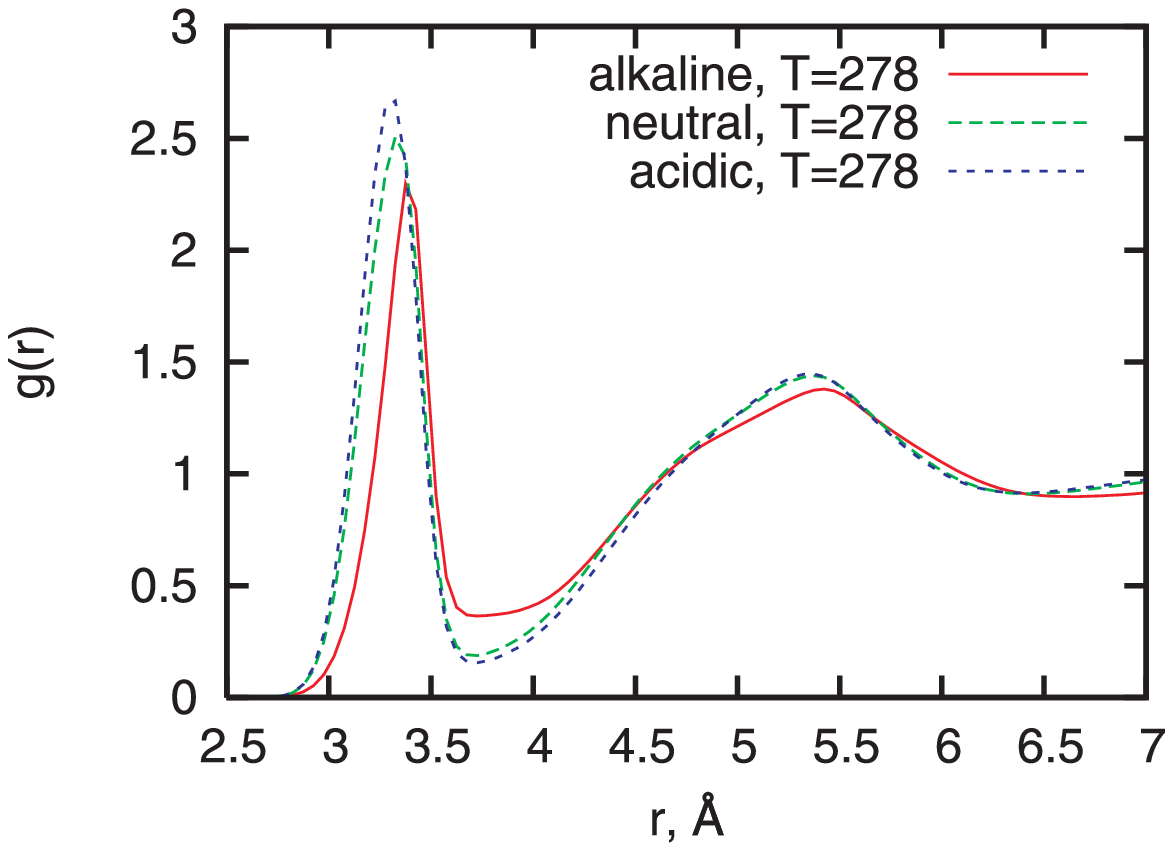}
\hfill
\includegraphics[width=0.48\textwidth]{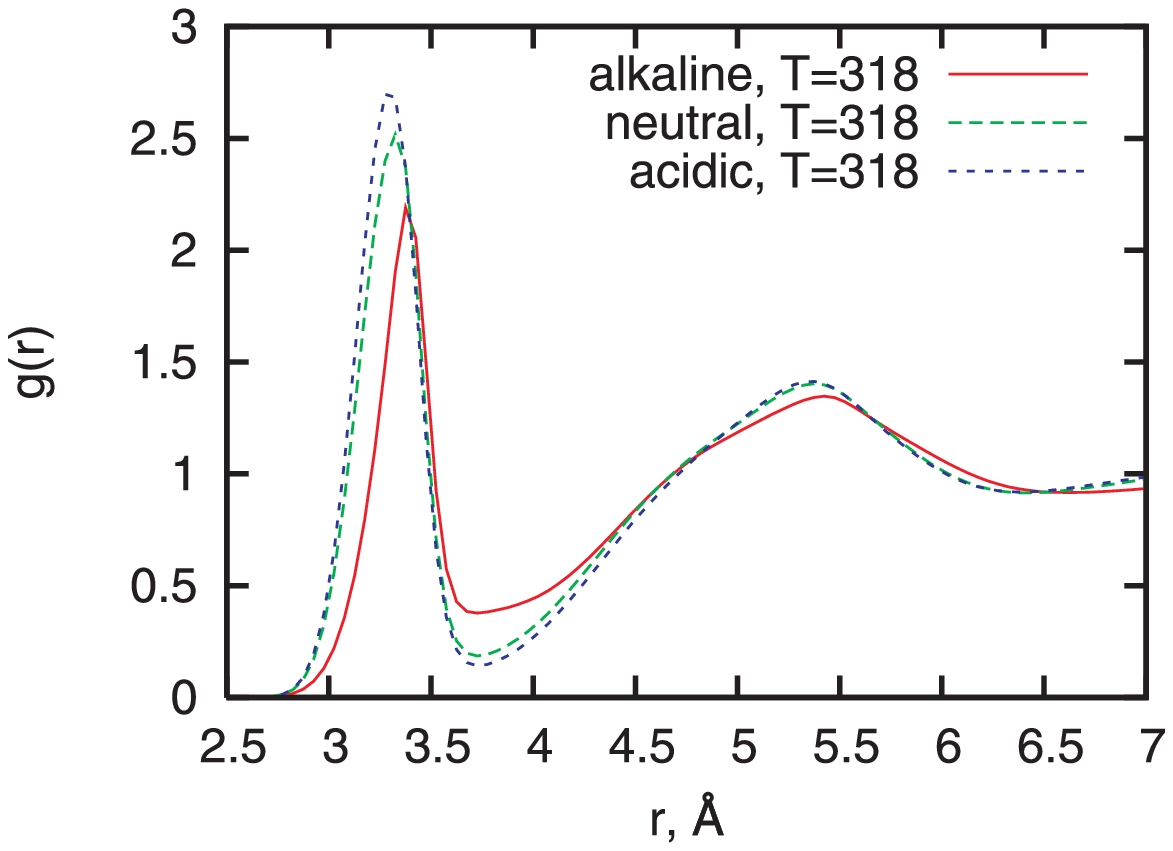}
}
\caption{(Color online) The uranium-hydrogen~(of water, OH$^-$ and H$^+$ ions) RDFs for
$T=278$~K~(left) and 318~K~(right). The color scheme is the same as in
figure~\ref{Fig:o-u}.}
\label{Fig:h-u}
\end{figure}

In figure~\ref{Fig:o-u} we present the RDFs for the uranium-oxygen correlation
for $T=278$~K (left hand panel) and 318~K (right hand panel).
The red solid lines denote the results for the alkaline, the green dashed lines --
for the neutral, and the blue dotted lines -- for the acidic solutions. If one
compares the RDFs between the two temperatures, it is clear that $T=278$~K case
is characterized by the higher first peaks, indicating more stable
formations under a lower temperature. One can also make another
apparent observation: for both considered temperatures, extra OH$^-$
ions in the solution favor the uranium-oxygen attraction while
the neutral and acidic conditions demonstrate roughly the same weaker
correlations. The roots of such a specific behavior in the
alkaline case originate from a competition between OH$^-$ ions and
the water molecules to be bound to the uranyl ions. Since the hydroxide ions
possess a negative charge, they are capable of pushing out the water molecules
from the uranyl shell, occupying the vacancies and becoming preferential neighbors
of the uranyls.
To illustrate this fact,
we show the uranium-hydrogen RDFs in figure~\ref{Fig:h-u}.
The distribution
belonging to the alkaline case demonstrates the lower first peak
reflecting a lack of hydrogens in the uranyl hydration shell in
comparison with the neutral and acidic solutions.
Another
consequence is the U--H peak shift toward larger distances since the O-H
axis in the OH$^-$ ions tends to be oriented directly from the uranium
in contrast to water molecules. An acidic case yields the U--H peaks higher than the neutral peak due to extra H$^+$ ions in the
bulk, preventing the uranyl ligands from being hydrolyzed.
A similar tendency to replace water molecules by OH$^-$ ions and to keep the
number of the ligands unchanged was previously found in \cite{NATO_uranyl}.

\begin{figure}[ht]
\centerline{
\includegraphics[width=0.45\textwidth]{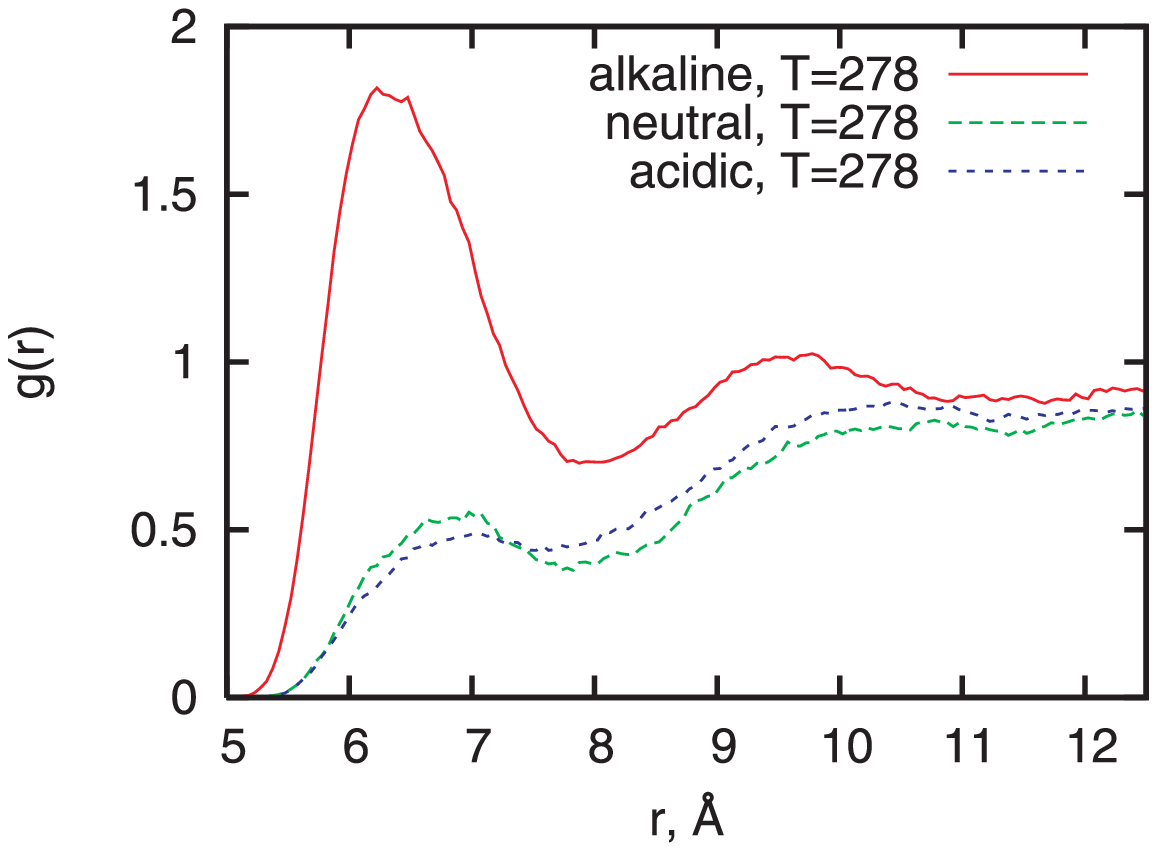}
\hfill
\includegraphics[width=0.45\textwidth]{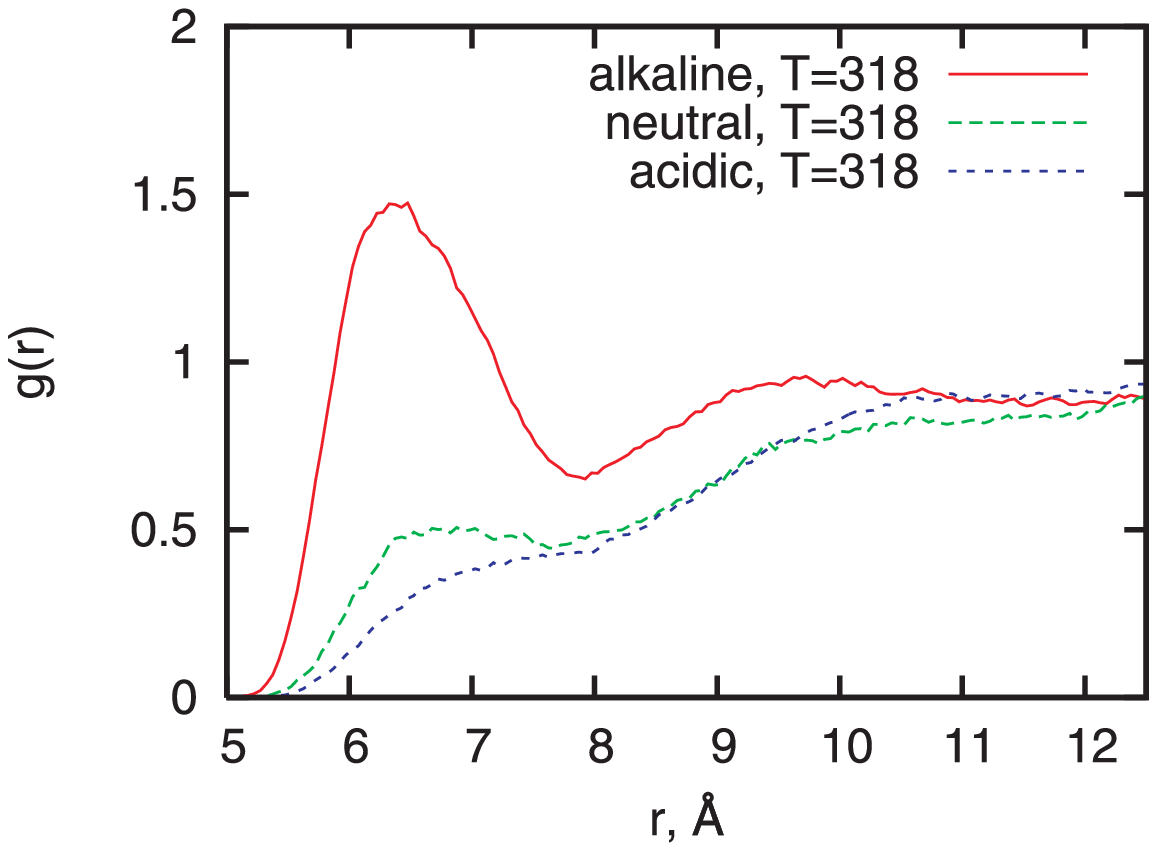}
}
\caption{(Color online) The uranium-uranium RDFs for $T=278$~K~(left) and 318~K~(right).
The color scheme is the same as in figure~\ref{Fig:o-u}.}
\label{Fig:u-u}
\end{figure}

Next we analyze the uranyl-uranyl coordination in terms of the uranium-uranium
and the uranium-oxygen (of uranyl) RDFs. In figure~\ref{Fig:u-u} the U--U
distributions for $T=278$~K and 318~K are presented.
The temperature effect follows the same pattern as the one observed for
the uranyl-water correlation. Also, the neutral and the acidic cases
behave similarly, while the alkaline one appears to differ from
these two. Contrary to the neutral and the acidic solutions, one
can see a prominent tendency for the uranyl-uranyl association in an
alkaline solution. Such an intense uranyl correlation (attraction)
in the alkaline solutions is a consequence of an increased screening of
the uranyls by OH$^-$ ions and the attraction between the uranyl and shells
(negatively charged by the hydroxide ions) of the other uranyls. That is why
the alkaline solution demonstrates the highest association ability in
spite of a weaker association in the neutral and the acidic solutions.

\begin{figure}
\centerline{
\includegraphics[width=0.45\textwidth]{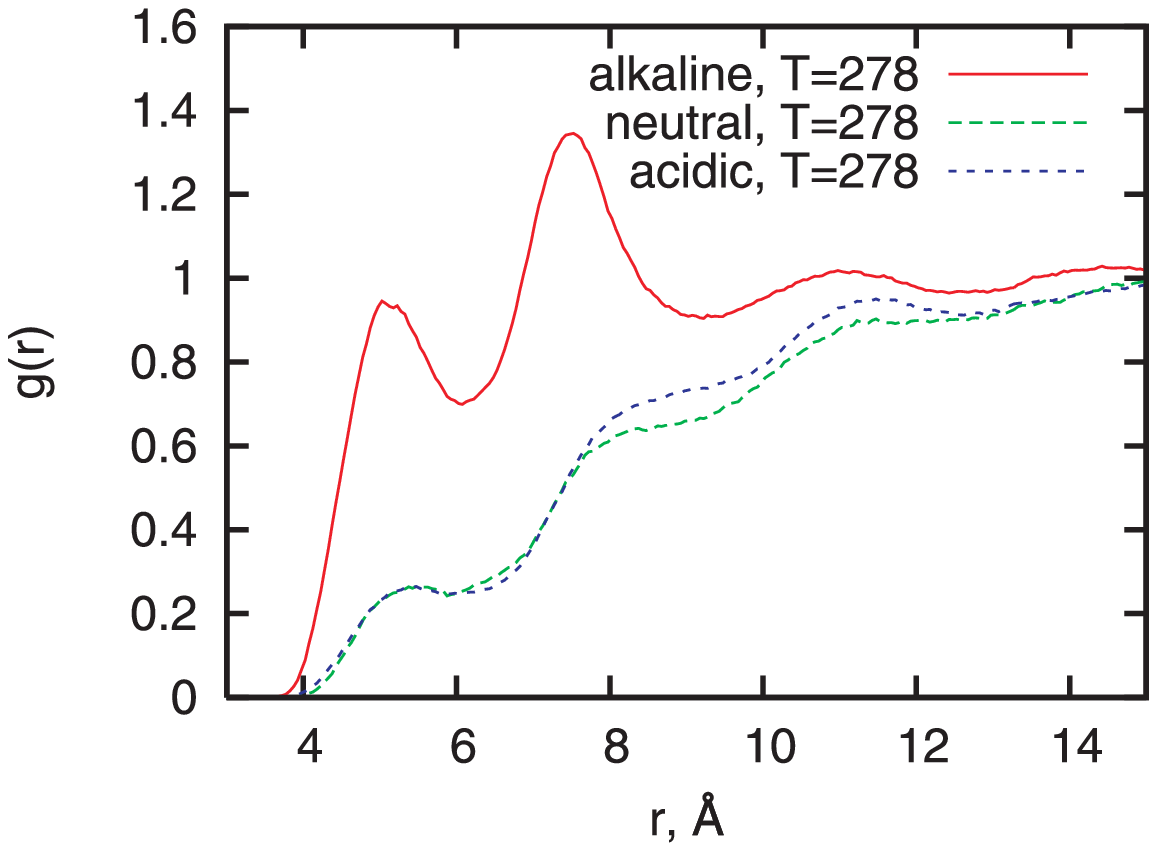}
\hfill
\includegraphics[width=0.45\textwidth]{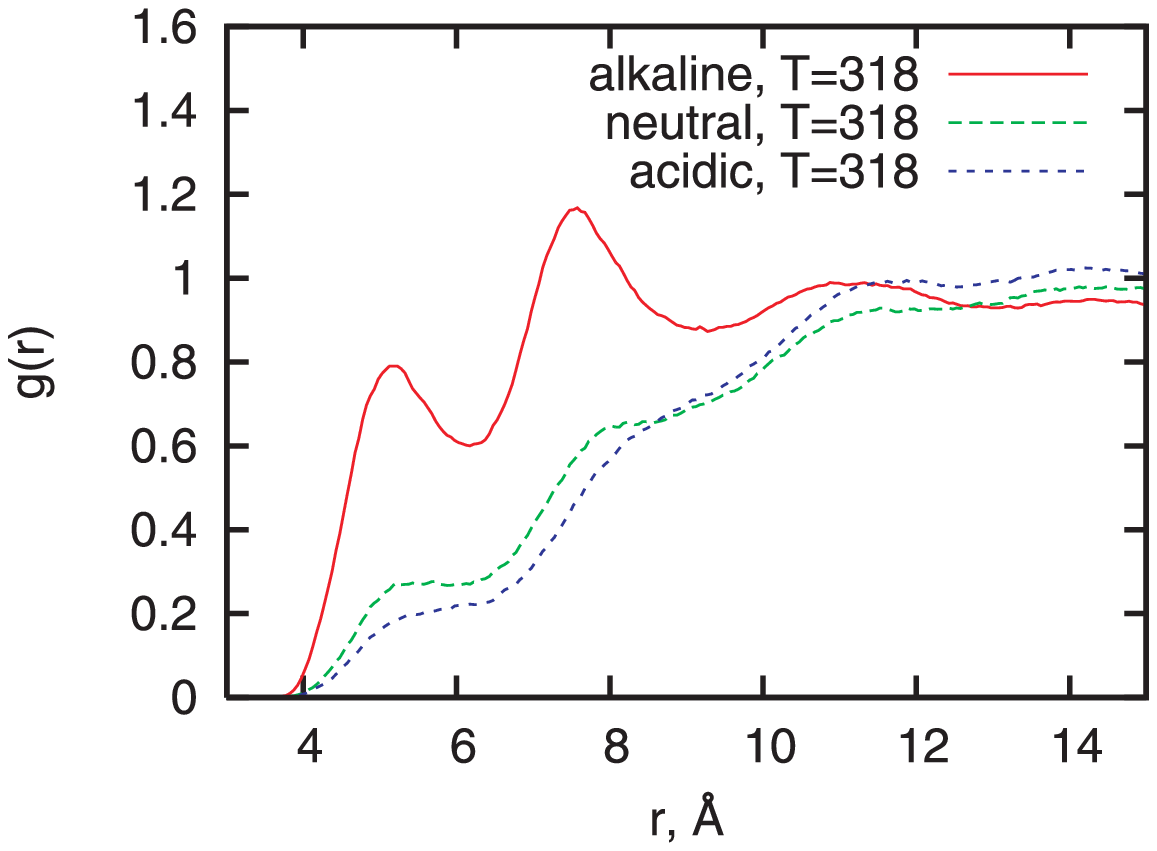}
}
\caption{(Color online) The uranium-oxygen (of uranyl) RDFs for $T=278$~K~(left) and 318~K~(right).
The color scheme is the same as in figure~\ref{Fig:o-u}. Note that the intramolecular
U${=}$O bonds do not contribute to the RDFs shown.}
\label{Fig:u-o1}
\end{figure}

Next we explore the correlation between the uranium and the oxygens of the
uranyls. In figure~\ref{Fig:u-o1} the corresponding RDFs for $T=278$~K (left hand
panel) and 318~K (right hand panel) are collected.
As one can expect, the RDFs for the alkaline solution indicate a stronger
correlation than the RDFs for the neutral and acidic cases. Nevertheless, all
these RDFs show the peaks near 4.5--5.5~{\AA} and 7--8~\AA. Temperature
produces a very similar effect by decreasing the peak heights.

One has to note that the RDFs in figures~\ref{Fig:u-u} and~\ref{Fig:u-o1}
are properly normalized and converge to unity at large distances but  a shorter distance range is chosen
in the plots to show the structural details.

\subsection{Uranyl-uranyl mutual orientation}

\begin{figure}[ht]
\centerline{
\includegraphics[width=0.4\textwidth]{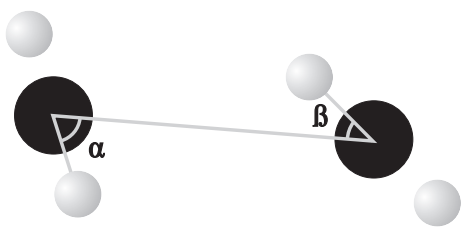}
}
\caption{The uranyl-uranyl mutual orientation and characteristic angles.}
\label{Fig:2U}
\end{figure}

\begin{figure}
\includegraphics[width=0.475\textwidth]{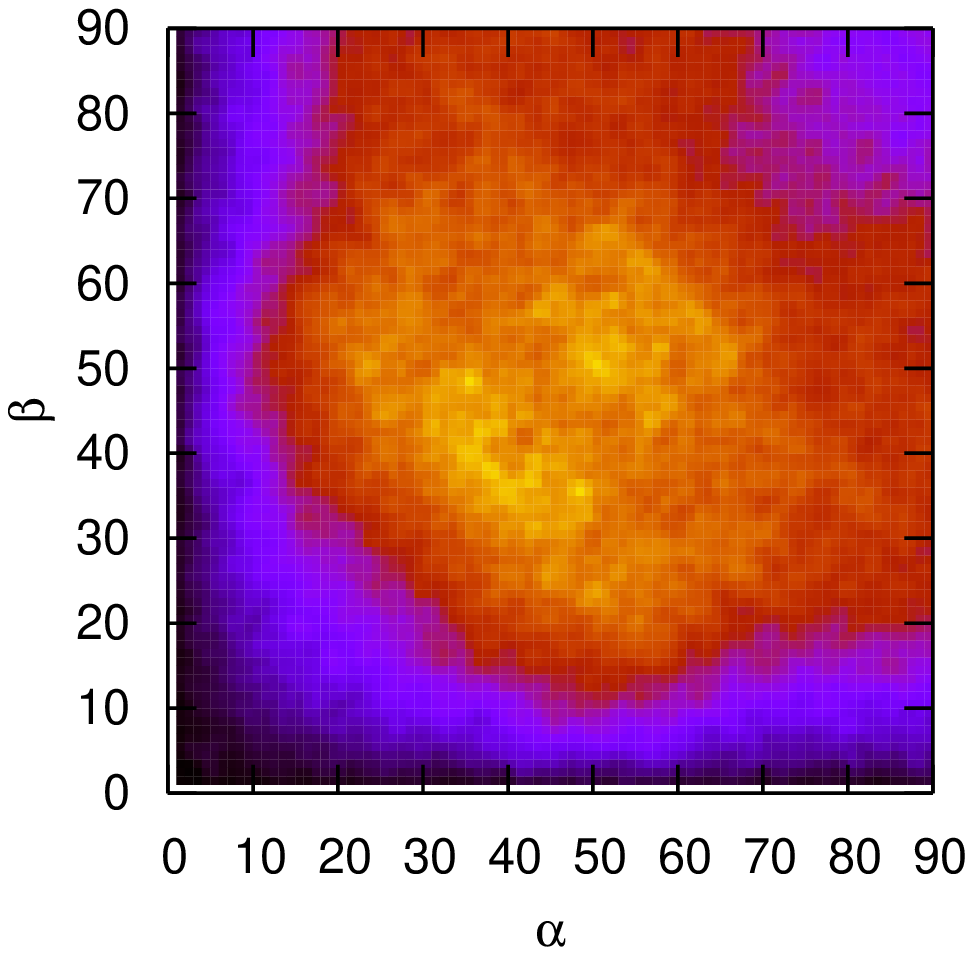} \hfill
\includegraphics[width=0.475\textwidth]{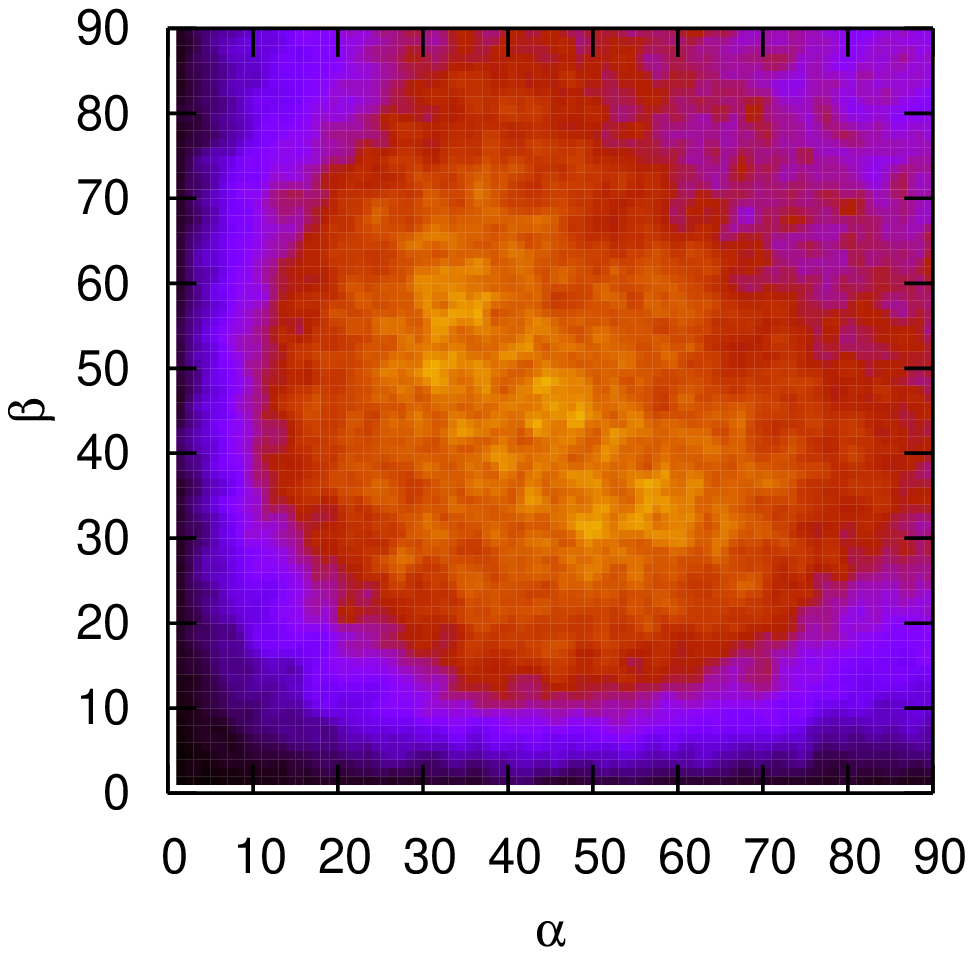} \\
\includegraphics[width=0.475\textwidth]{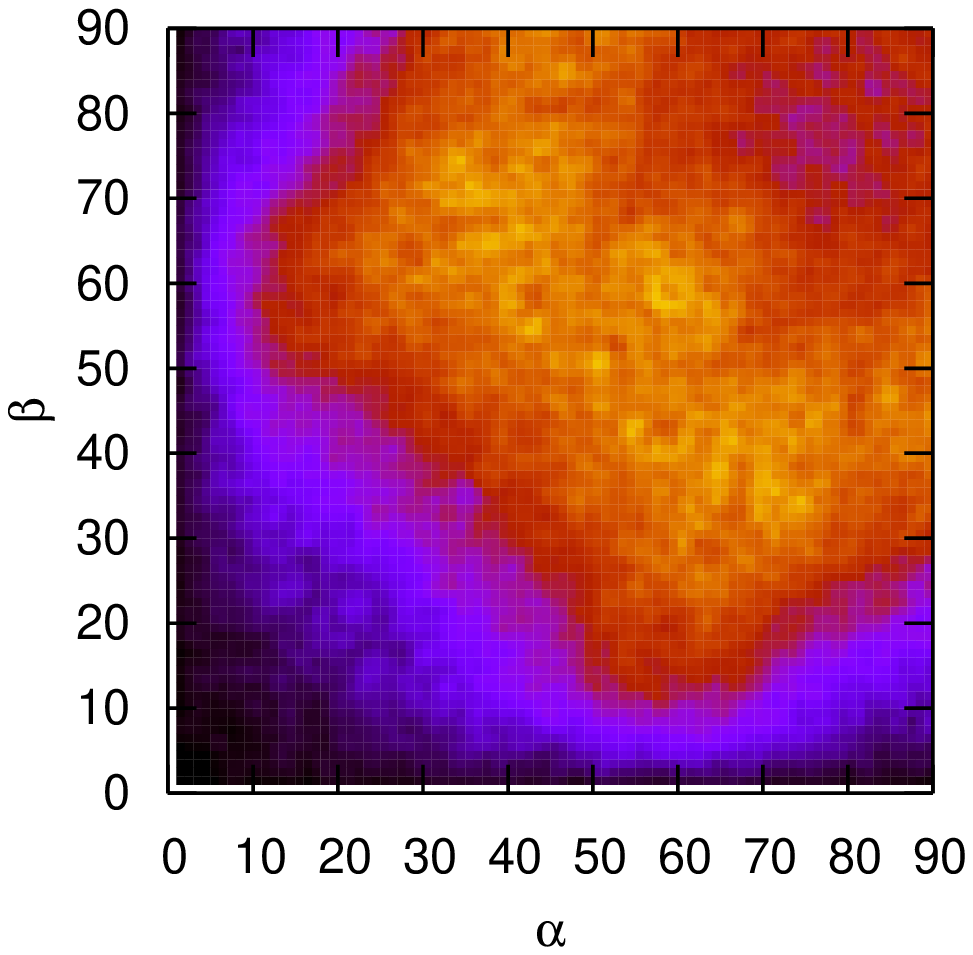} \hfill
\includegraphics[width=0.475\textwidth]{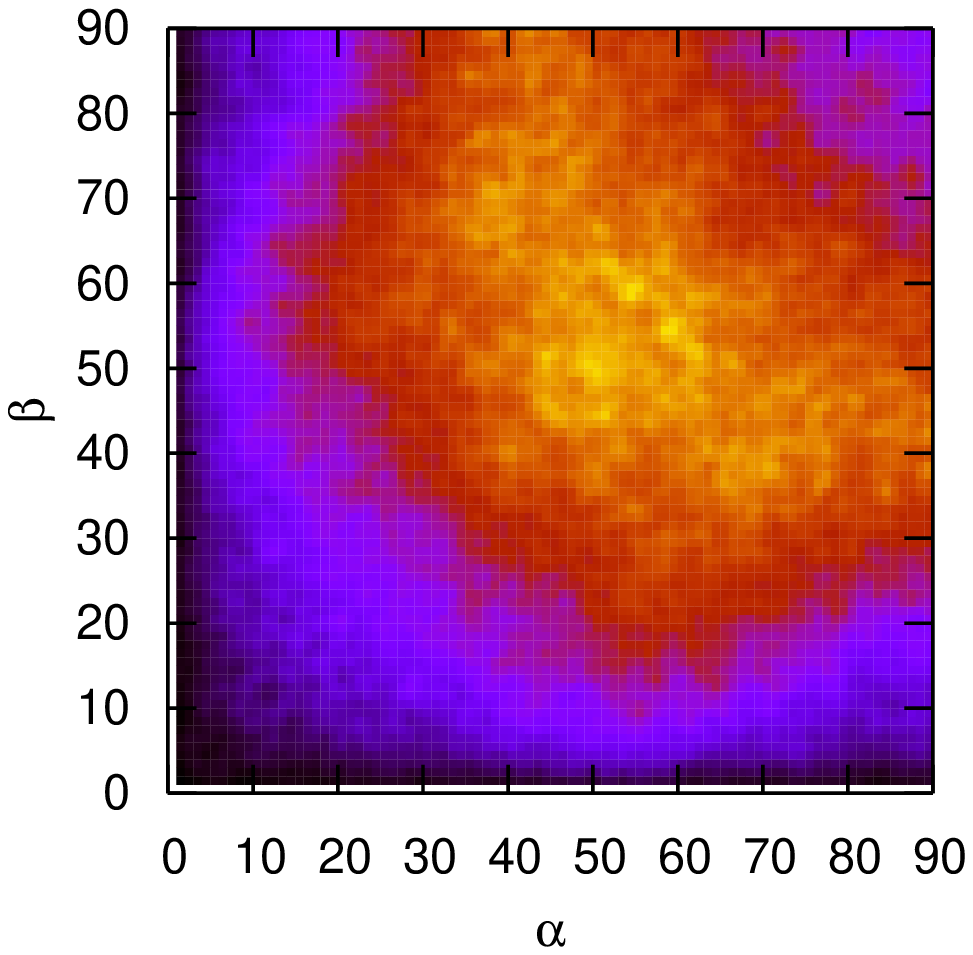} \\
\includegraphics[width=0.475\textwidth]{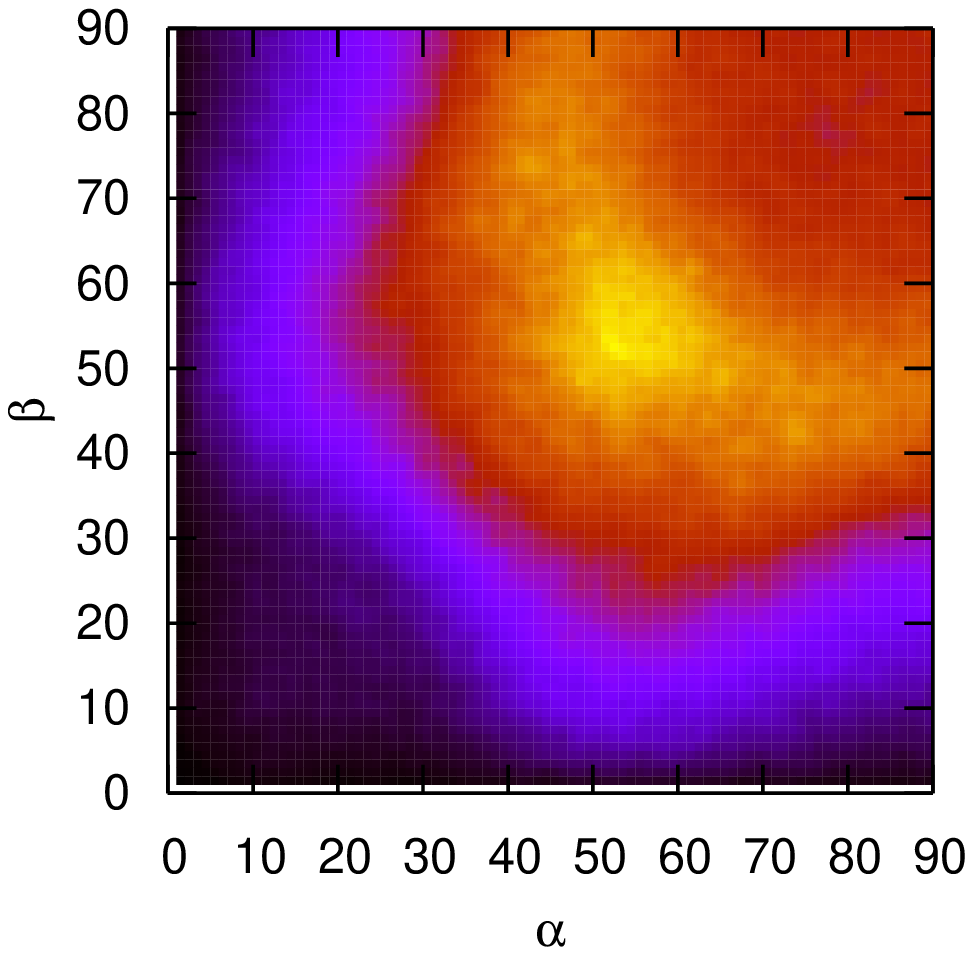} \hfill
\includegraphics[width=0.475\textwidth]{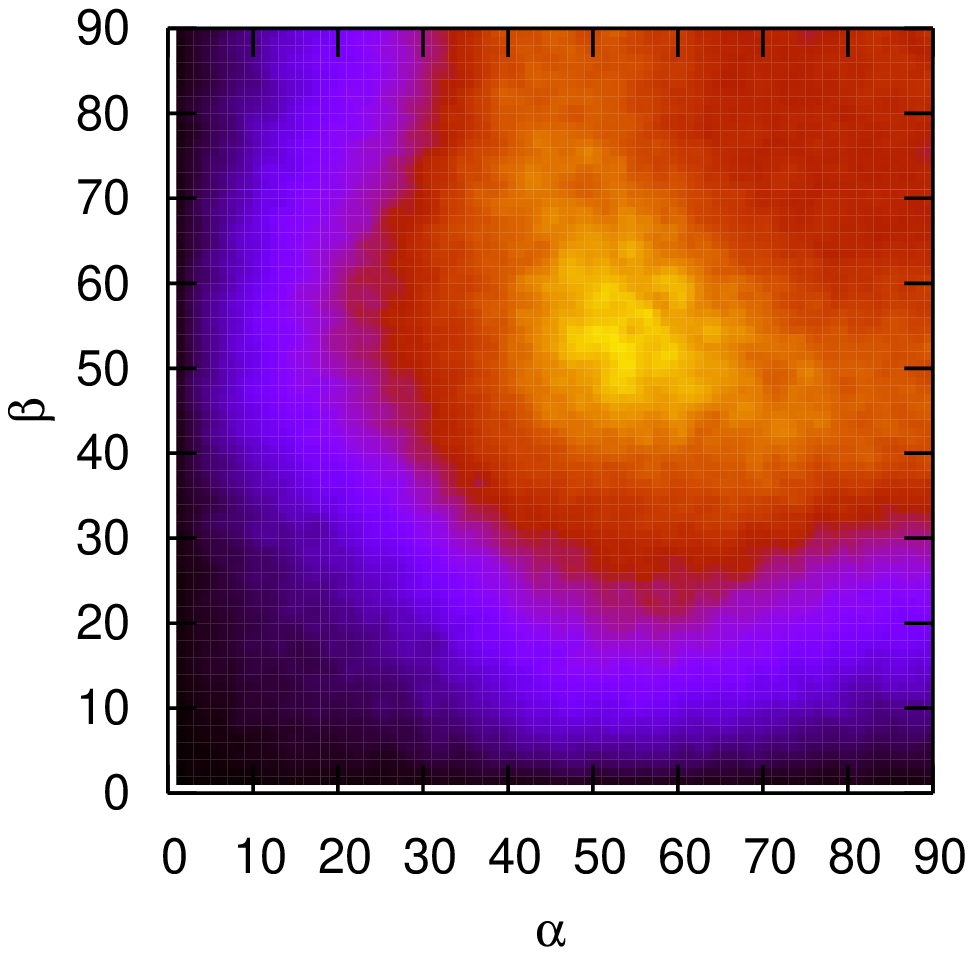}
\caption{(Color online) The uranyl-uranyl angular distributions for the acidic~(top),
neutral~(center) and alkaline~(bottom row) solutions at $T=278$~K~(left) and
318~K~(right hand column). The colors correspond to the probability
from low~(dark) to high~(bright) to find a corresponding configuration.
Color scheme is scaled in such a way that the maximum values of all
the distributions are of the same intensity.}
\label{Fig:Distr}
\end{figure}

We made the angular analysis in order to clarify how
the uranyls are mutually oriented being involved in the associates.
During the simulation, the coordinates of the uranyls were stored and then
processed. A straight and simple criterion to test whether the uranyls
are forming an associate is the distance between uraniums: the
location of the first minima of the U--U RDFs (figure~\ref{Fig:u-u}) can
be used for this purpose. Since the minima locations are smeared
for different temperatures, a single distance threshold
$r=7.5$~{\AA} was utilized.
The angular analysis protocol is as follows: at every simulation step,
one can assign a vector along the uranyl axis O${=}$U${=}$O. Next we consider a
vector between two uranium ions when the distance between them is less
than 7.5~{\AA}. A description of the mutual uranyl orientations can be treated
in terms of two angles $\alpha$, $\beta$ between each of O${=}$U${=}$O axis and
the U--U vector (see figure~\ref{Fig:2U}). The corresponding symmetrized
probability distributions are shown in figure~\ref{Fig:Distr}.
As one can see from figure~\ref{Fig:Distr}, the angular distributions
show a pronounced dependence on pH of the solution, while the temperature
effect is less apparent. We start with the acidic
solutions: a bright spot in the symmetric range 30--65$^\circ$
indicates a configuration when both uranyls are
tilted to the vector connecting the uraniums. A wide range
of available angles speaks in favor of a ``breathing''
behavior of such associates. For a neutral
solution at $T=278$~K, one can discriminate two probability spots
at $\alpha={}$25--45$^\circ$, $\beta={}$55--75$^\circ$ and vice versa. At 318~K,
the distribution demonstrates the similar shape though
slightly smeared. The alkaline distribution
for 278~K shows the symmetric spot at $\alpha=\beta={}$45--60$^\circ$ with a
tiny possibility for configurations with both angles equal to $90^\circ$
(parallel orientation of uranyls). Similarly to the neutral case, the temperature
makes a smearing effect when going to 318~K. One can see that the
distributions for the alkaline solutions are characterized by more strict
and localized spots. This is due to the hydroxide ions in the uranyl
hydration shells when more electrostatic players take part in the
stabilization of the polynuclear associate.

All the distributions indicate no associates with the arrangement of
uranyls on a straight line (both angles are $0^\circ$), and no
associates with the perpendicular orientation (one angle is 0$^\circ$ and
the other -- $90^\circ$).

\subsection{Lifetimes and associate fractions}

Apart from the structural information on the ordering of the uranyls in an
associate, it is also in\-te\-res\-ting to measure the average time spent by
them in the associated state. To obtain an information on the
lifetimes of the uranyl associates, we monitor their trajectories
during the simulation. Our strategy is: for any
observed dimer (the distance between two uraniums is less than
7.5~\AA) we start its individual ``stopwatch'' and let the
stopwatch go during the time when the distance criterion remains
satisfied. When an uranyl leaves an associate, the stopwatch is stopped.
The actual procedure involves several steps: first (i), we determine the atom
distributions and identify the uranyls involved in the associates
($t=0$), next (ii) we repeat this procedure (checking the
configuration) every 5 elementary steps, that is on 1.0 fs
interval. (iii) When the associate is broken ($t = \tau$), we
update the lifetime distribution $f(\tau)$. If another uranyl attaches to
a dimer (the distance to any of dimer uraniums is less than 7.5),
this associate becomes a trimer and a corresponding ``stopwatch''
is ascribed.

Next, based on the lifetime distribution $f(\tau)$, we can
estimate the average lifetimes as:
\begin{equation}
\langle\tau\rangle = {\frac{\int_0^{\tau_{\mathrm{max}}} \tau
f(\tau) \rd\tau} {\int_0^{\tau_{\mathrm{max}}} f(\tau) \rd\tau}}\,,
\end{equation}
where $\tau_{\mathrm{max}}={}$200~ps is taken as the integration limit~($\infty$)
since the distributions $f(\tau)$ have completely decayed at these times.

In table~\ref{LTs}, we present the average lifetimes for the uranyl associates
(dimers and trimers) at $T=278$ and 318~K. One can see that a decrease
of the temperature extends the average lifetimes $\langle\tau\rangle$.
Another factor favoring a formation of the associates is the presence of OH$^-$
ions. That is why the strongest association uranyl capability is found for the
alkaline solution at 278~K. As it was shown above, the association capability
in the alkaline case is sufficiently increased due to the uranyl screening
by the hydroxide ions. In the absence of OH$^-$ ions, the acidic and neutral
solutions behave in a similar way demonstrating a relatively close results
within the uncertainties. Due to this, the lifetime for the acidic case at
278~K is longer than the one for the neutral case at 278~K. The average trimer
lifetimes for the acidic case at 278 and 318~K do not fit the overall
temperature tendency but the differences originate from the relatively short
association events and do not change the main trends.

\begin{table}[h]
\caption{The average lifetimes of the uranyl associates. Time units are -- ps.}
\label{LTs} \vspace{2ex}
\begin{center}
\begin{tabular}{|c|c|c|c|c|}
\hline
\raisebox{-1.7ex}[0pt][0pt]{}
& \multicolumn{2}{|c|}{dimers} & \multicolumn{2}{|c|}{trimers}  \\
\cline{2-5}
         & 278 K & 318 K & 278 K & 318 K\\
\hline\hline
acidic   & 14.18 & 5.71 & 1.84 & 2.55\\
\hline
neutral  & 13.79 & 6.21 & 3.15 & 2.65\\
\hline
alkaline & 20.98 & 10.57 & 10.14 & 5.00\\
\hline
\end{tabular}
\end{center}
\end{table}

Sometimes \cite{Spohr97,Laage08} in the studies of the association lifetimes, besides the
distance criterion, the authors introduce an additional tolerance time.
During this time, an associate is allowed to be broken and to be united
back so that the lifetime ``stopwatch'' keeps running. The purpose of
such a deployment is to avoid an overestimation of the shorter lifetimes
due to sequent broken/united events. In our study we did not
utilize this formalism because it is not well defined and does not
affect the general conclusions.

Other quantities to consider are the fractions of the uranyls taking part
in the formation of dimers and trimers. These fractions are
defined as the number of the uranyls in the corresponding type of associates
divided by the total number of the uranyls in the solution. The criteria for
dimer and trimer creation events are the same as before. In table~\ref{As} the
fractions are collected. As one can expect, the same tendencies
can be found here as above for the average lifetimes: the low temperature
and the presence of OH$^-$ ions are the factors increasing the capability
of the uranyls to associate.
\begin{table}[h]
\caption{The fractions of the uranyls involved in the associates.}
\label{As} \vspace{2ex}
\begin{center}
\begin{tabular}{|c|c|c|c|c|}
\hline
\raisebox{-1.7ex}[0pt][0pt]{}
& \multicolumn{2}{|c|}{dimers} & \multicolumn{2}{|c|}{trimers}  \\
\cline{2-5}
& 278 K & 318 K & 278 K & 318 K\\
\hline\hline
acidic  & 0.132 & 0.077 & 0.008 & 0.006\\
\hline
neutral  & 0.154 & 0.133 & 0.019 & 0.023\\
\hline
alkaline  & 0.411 & 0.350 & 0.218 & 0.145\\
\hline
\end{tabular}
\end{center}
\end{table}

Despite the similar conclusions drawn from the lifetimes and the fractions,
one has to note that the average lifetime does not contain the same information
as the fraction of the associated uranyls: the same fraction of the associated
uranyls can be obtained as a result of many short or,
alternatively, fewer but longer (in time) formations.

\section{Conclusions}

The purpose of this study is to clarify in what way the temperature and pH
effect the formation of the polynuclear associates in the uranyl aqueous
solutions. For this purpose, we performed a series of MD simulations of the uranyl
aqueous solutions at temperatures 278 and 318~K. To model different pH's
the neutral system (containing waters and uranyls only) was mixed with
the additional H$^+$ or OH$^-$ ions mi\-mi\-cking an acidic, or alkaline
solutions. During the simulations we collected the necessary statistics on the radial
distributions functions describing the uranyl hydration shells and
the uranyl-uranyl correlation. Comparing the RDFs at the temperatures
278 and 318~K one can conclude that the increase of temperature
leads to a decrease of the correlation (the peaks of the RDFs become lower).
It is also found that in the acidic and the neutral solutions, the uranyl
hydration shells stay unchanged, while in the alkaline case, the uranyls are
preferably hydrated by the hydroxide ions. The modified hydration in the
alkaline case results in the increased uranyl attraction
contrary to a weak uranyl correlation in the neutral and the acidic solutions.
The negatively charged OH$^-$ ions screen the UO$_2^{2+}$ ions allowing them
to form more stable associates. In addition to the RDFs, we
introduced the angular distributions reflecting the mutual orientation of
the uranyls involved in the polynuclear associates. A temperature
effect on the angular distributions is less apparent than the effect made by
pH: indeed one can see that the distributions change the shape
from smeared to a strict one when going from the acidic to the neutral and
then to the alkaline case. This is due to the presence of the hydroxide ions in
the uranyl hydration shells, that stabilize the uranyl-uranyl associates.
For every solution we calculated the fraction of the uranyls taking
part in the associates. Besides the fractions we also monitored the lifetimes
of the uranyl dimers and trimers, built corresponding lifetime
distributions, and extracted the average lifetimes for each case. The
conclusions drawn agree with the ones from the RDFs: the low temperature and
the alkaline environment are the factors favoring the long-lasting uranyl
associates.

\section*{Acknowledgements}
The MD calculations were performed on the clusters of Ukrainian Academic Grid.

\ukrainianpart

\title{Вплив температури та рівня рН на асоціативність у водних розчинах уранілу}
\author{М.~Дручок, М.~Головко}
\address{
Інститут фізики конденсованих систем НАН України,
вул. Свєнціцького 1, Львів 79011, Україна
}

\makeukrtitle

\begin{abstract}
\tolerance=3000%
Проведено дослідження процесів асоціативності у водних розчинах
уранілу. Для цього проведено низку моделювань методом молекулярної
динаміки. Під час моделювання проведено моніторинг фракцій іонів
уранілу, що беруть участь у формуванні димерів та тримерів. Також
зібрано розподіли імовірності за часами життя уранілових димерів
та тримерів. Розглянуто вплив двох факторів -- температури та рівня
рН середовища~-- на здатність іонів уранілу до формування асоціатів.
Виявлено, що збільшення температури знижує асоціативність, при цьому також
скорочуються часи життя асоціатів і зменшуються фракції іонів
уранілу у асоціатах. Вплив рівня рН середовища змодельовано додаванням
іонів H$^+$ чи OH$^-$ до ``нейтрального'' розчину. Виявлено, що
наявність у розчині іонів OH$^-$ є сприятливим чинником для
асоціативності, в той час як наявність іонів H$^+$ призводить
до протилежного ефекту. Для усіх розчинів проведено конфігураційний
аналіз взаємної орієнтації іонів уранілу, що перебувають у
асоціативному стані.
\keywords молекулярна динаміка, водний розчин уранілу,
асоціативність, pH, температура

\end{abstract}

\end{document}